\documentclass{elsart}

\usepackage{amsmath,amssymb,graphicx,bm}

\newcommand{\beqn}{\begin{equation}}
\newcommand{\eeqn}{\end{equation}}
\newcommand{\bea}{\begin{eqnarray}}
\newcommand{\eea}{\end{eqnarray}}

\newcommand{\nexp}{n}

\newcommand{\fmi}{\, \text{fm}^{-1}}

\newcommand{\wt}{\widetilde}
\newcommand{\openone}{\leavevmode\hbox{\small1\normalsize\kern-.33em1}}

\newcommand{\Hzero}{T_{\rm rel}}
\newcommand{\flow}{s}

\begin{document}

\begin{frontmatter}

\title{Decoupling in the Similarity Renormalization Group 
          for Nucleon-Nucleon Forces}

\author[OSU]{E.D.\ Jurgenson},
\ead{jurgenson.1@osu.edu}
\author[OSU,MSU]{S.K.\ Bogner},
\ead{bogner@nscl.msu.edu}
\author[OSU]{R.J.\ Furnstahl},
\ead{furnstahl.1@osu.edu}
\author[OSU]{R.J.\ Perry}
\ead{perry@mps.ohio-state.edu}
\address[OSU]{Department of Physics, The Ohio State University, Columbus, OH 43210}
\address[MSU]{National Superconducting Cyclotron Laboratory and Department 
  of Physics and Astronomy, Michigan State University, East Lansing, MI 48844}


\begin{abstract}
Decoupling via the Similarity Renormalization Group (SRG) of low-energy nuclear 
physics from high-energy details of the nucleon-nucleon interaction is examined 
for two-body observables and few-body binding energies. 
The universal nature of 
this decoupling is illustrated and 
errors from suppressing high-momentum modes above the decoupling 
scale are shown to be perturbatively small.

\end{abstract}

\end{frontmatter}
\maketitle

\section{Introduction}
\label{sec:intro}

The Similarity Renormalization Group
(SRG)~\cite{Glazek:1993rc,Wegner:1994,Kehrein:2006}  provides a compelling
method for evolving internucleon forces to softer 
forms~\cite{Bogner:2006srg,Bogner:2007srg}. While observables are unchanged
by the  SRG's unitary transformations, 
the contributions from high-momentum intermediate states 
to low-energy observables is modified by the running transformation. 
In particular, the SRG
as implemented in  Refs.~\cite{Bogner:2006srg,Bogner:2007srg} has the effect of
partially  diagonalizing the momentum-space potential to a width of order the
evolution parameter.  Because of this partial diagonalization, 
one anticipates a direct
decoupling of low-energy observables from high-energy degrees of
freedom. 

In Ref.~\cite{Bogner:2007srg}, evidence for decoupling at low momentum was
shown  for the Argonne V18 \cite{Wiringa:1994wb} potential in calculations of
phase shifts and the deuteron. 
In this letter, we extend the demonstration of
decoupling to nucleon-nucleon  (NN) potentials from chiral effective field
theory (EFT)~\cite{N3LO,N3LOEGM} and to few-body 
nuclei up to $A = 6$ to verify its
universal nature and to show quantitatively that the residual coupling is
perturbative above the energy corresponding to the SRG  evolution parameter. 
One might imagine that there is little to gain by evolving already-soft 
chiral EFT potentials to lower momentum, but recent work  shows
significant advantages for the nuclear few- 
and many-body problem~\cite{Bogner:2007rx}.
This is consistent with the additional decoupling
we find here.

The practical test for decoupling is whether changing high-momentum matrix
elements of the potential changes low-energy observables.
Our strategy is to first evolve the initial potential $V_{\rm NN}$ 
with the SRG equations to obtain the SRG potential $V_\flow$,
where $s$ denotes the flow parameter of the
transformation. 
Then we apply a parametrized
regulator to cut off the high-momentum part of the evolved 
potential in a controlled way. 
This cutoff potential is used to calculate
few-body  observables and their relative errors. By varying the parameters of
the regulator  and correlating them with errors in the calculated
observables, we have a
diagnostic tool to quantitatively analyze the decoupling.

In Section~\ref{sec:methods}, we give background on decoupling in the
SRG and the use of a smooth regulator to
quantify decoupling. 
In  Section~\ref{sec:phase_shifts}, we analyze the relative error
between phase shifts calculated with uncut and cut potentials 
and find a clean power law dependence at
larger momenta that follows from  perturbation theory. 
In Section~\ref{sec:deuteron}, 
we examine observables for the deuteron bound state
and find the same perturbative behavior for the errors.  
Using the No-Core Shell
Model (NCSM)~\cite{NCSMC12,NCSM2,NCSM3,Nogga:2005hp,Navratil:2007we},
we verify this behavior for few-body calculations in
Section~\ref{sec:ncsm_calcs}
and summarize in Section~\ref{sec:summary}.

\section{Decoupling in the SRG}
\label{sec:methods}

As in Ref.~\cite{Bogner:2006srg}, 
we apply the similarity renormalization group 
(SRG) transformations to NN interactions based on the flow equation formalism 
of Wegner~\cite{Wegner:1994}. The evolution or flow of the Hamiltonian with a 
parameter $\flow$ is a series of unitary transformations,
\beqn
H_\flow = U_\flow H U^\dagger_\flow \equiv \Hzero + V_\flow \;,
\label{eq:Hflow}
\eeqn
where $\Hzero$ is the relative kinetic energy and $H = \Hzero + V$ is the
initial  Hamiltonian in the center-of-mass system. Equation (\ref{eq:Hflow})
defines the  evolved potential $V_\flow$, with $\Hzero$ taken to be
independent of $\flow$.  Then $H_\flow$ evolves according to
\beqn
\frac{dH_\flow}{d\flow} = [\eta_\flow,H_\flow] \;,
\eeqn
with
\beqn
\eta_\flow = \frac{dU_\flow}{d\flow} U^\dagger_\flow 
= -\eta^\dagger_\flow \;.
\eeqn
Choosing $\eta_\flow$ specifies the transformation, which is taken as the 
commutator of an operator, $G_\flow$, with the Hamiltonian, 
\beqn
\eta(\flow) = [G_\flow, H_\flow] \;,
\eeqn
so that
\beqn
\frac{dH_\flow}{d\flow} = [ [G_\flow, H_\flow], H_\flow] \;.
\label{eq:commutator}
\eeqn
Applications to nuclear physics to date in a partial-wave
momentum basis have used 
$G_\flow = T_{\rm rel}$~\cite{Bogner:2006srg}, but one could
also use momentum-diagonal operators such as 
$\Hzero^2$, $\Hzero^3$, or the running diagonal Hamiltonian
$H_D$, as advocated by Wegner~\cite{Wegner:1994}. 

The source of decoupling is the partial diagonalization of the Hamiltonian
by the SRG evolution.  
For the NN interaction, the flow equation
(\ref{eq:commutator}) can be simply evaluated in the space of 
discretized relative
momentum NN states~\cite{Bogner:2006srg}.  
For a given partial wave, with units where $\hbar^2/M = 1$,
we define diagonal matrix elements of momentum $k$ as
\beqn
   \langle k | H_\flow | k \rangle = \langle k | H_D | k \rangle
   \equiv e_k
   \;,  
\eeqn 
and
\beqn
   \langle k | T_{\rm rel} | k \rangle 
   \equiv \epsilon_k = k^2 \;.
\eeqn 
With Wegner's choice, $G_\flow = H_D$, the flow equation 
for each matrix element is
\beqn
  \frac{d}{ds} \langle k|H_\flow|k'\rangle =
  \sum_q (e_k + e_{k'} - 2 e_q) \langle k|H_\flow|q\rangle
      \langle q|H_\flow|k'\rangle
      \;.  \label{eq:mecommutator}
\eeqn
By considering
the trace of $H_\flow^2$, one can show 
that off-diagonal matrix elements as a whole decrease in magnitude
unless there are energy degeneracies~\cite{Wegner:1994}.  

If we take
$G_\flow = T_{\rm rel}$,
the flow equation for each matrix element is 
\bea
  \frac{d}{ds} \langle k|H_\flow|k'\rangle &=&
  \sum_q (\epsilon_k + \epsilon_{k'} - 2 \epsilon_q) \langle k|H_\flow|q\rangle
      \langle q|H_\flow|k'\rangle
      \nonumber \\
    &=& 
     -(\epsilon_k - \epsilon_{k'})^2 \langle k | V_s | k' \rangle
    \nonumber \\ &&  \null \quad +      
  \sum_q (\epsilon_k + \epsilon_{k'} - 2 \epsilon_q) \langle k|V_\flow|q\rangle
      \langle q|V_\flow|k'\rangle
      \;.  \label{eq:mecommutatorT}
\eea
In this case, it can be shown that off-diagonal
elements are not guaranteed to decrease monotonically
if $e_k - e_{k'}$ and $\epsilon_k - \epsilon_{k'}$
have opposite signs (see Ref.~\cite{Glazek:2007} for details).
However, this does not happen in the range of $s$ that has been
considered in the nuclear case because
of the dominance of the kinetic energy.

\begin{figure}[th]
\begin{center}
\includegraphics*[width=6.7cm]{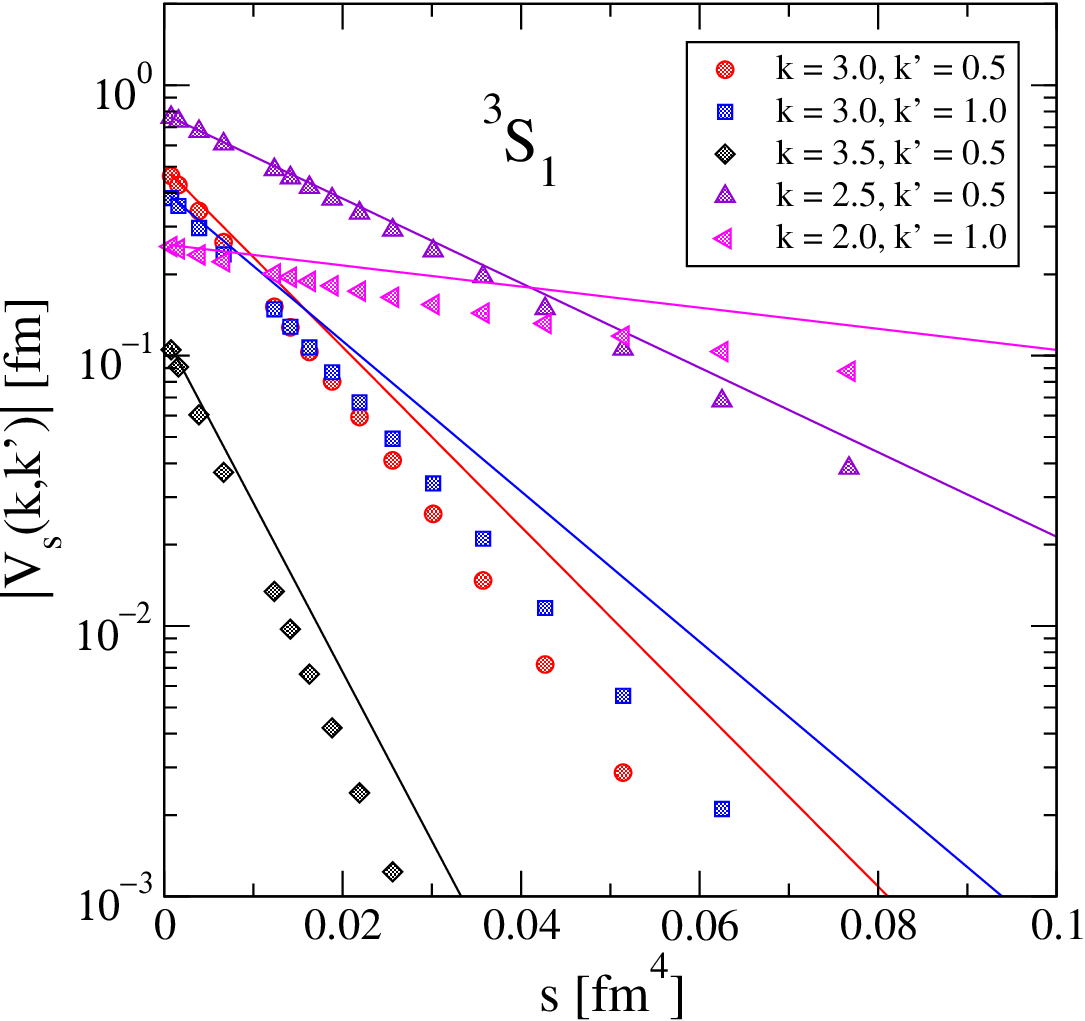}
\hspace*{.1in}
\includegraphics*[width=6.7cm]{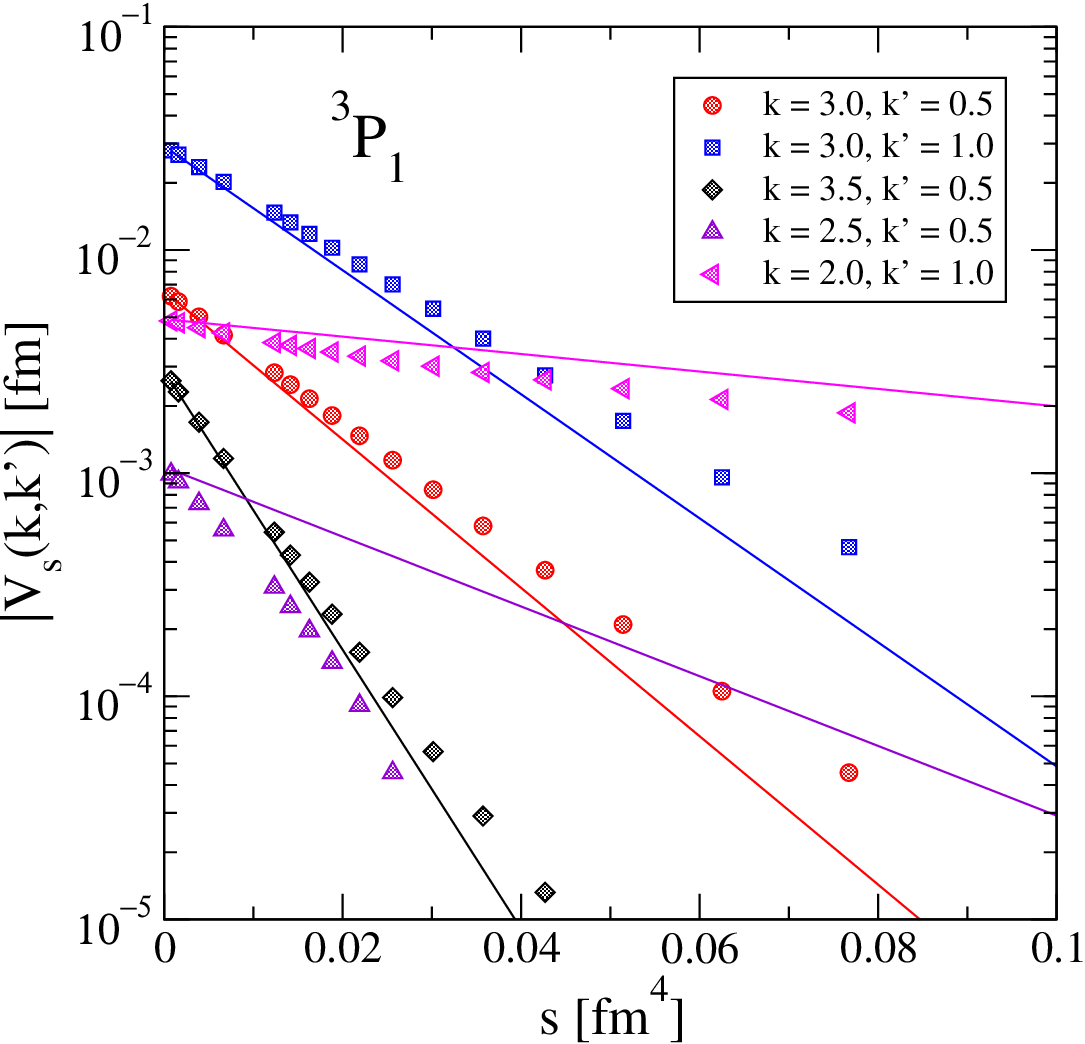}  
\end{center}

\caption{Absolute value of 
the matrix element $\langle k | V_\flow | k' \rangle $ for a
representative sampling of off-diagonal $(k,k')$ pairs as a function
of $s$, compared with the simple solutions from Eq.~(\ref{eq:exponential}),
which are straight lines (they agree at $s=0$).  Two partial waves are shown.
}
\label{fig:flowtest}
\end{figure}

This dominance can be used to find a semi-quantitative 
approximation for the flow of off-diagonal matrix elements by
keeping only the first term on the right side of Eq.~(\ref{eq:mecommutatorT}) 
(or, equivalently, substituting $\langle k | H_\flow | q \rangle
\rightarrow \epsilon_k\, \delta_{kq}$).
Then the flow equation applied to individual
off-diagonal matrix elements simplifies to
\beqn 
  \frac{d}{ds} \langle k|H_\flow|k'\rangle =
  \frac{d}{d\flow} \langle k | V_\flow | k' \rangle 
  \approx - (\epsilon_k - \epsilon_{k'})^2 \, \langle k | V_\flow | k' \rangle
  \;,
\eeqn
which has the simple exponential solution
\beqn
\langle k | V_\flow | k' \rangle \approx \langle k | V_{s=0} | k' 
       \rangle e^{-\flow(\epsilon_k - \epsilon_{k'})^2}\;.
       \label{eq:exponential}
\eeqn
In Fig.~\ref{fig:flowtest}, we plot 
$\langle k | V_\flow | k' \rangle \approx \langle k | V_{s=0} | k' \rangle$
and the approximation from Eq.~(\ref{eq:exponential}) versus $s$
for some representative
off-diagonal points in two partial waves.
In almost all cases the approximation gives a reasonable estimate
of the monotonic decrease to zero; in the one exception there
is a significantly more rapid decrease than predicted.

It is convenient to switch
to the flow variable $\lambda = 1/s^{1/4}$, which has
units of $\fmi$, since Eq.~(\ref{eq:exponential})
shows that $\lambda$ is a measure of the resulting
diagonal width of $V_\flow$ in momentum space. 
More precisely, the matrix $\langle k | V_\flow | k' \rangle $ 
plotted as a function of kinetic energies $k^2$ and $k'{}^2$
will rapidly go to zero outside of a diagonal band roughly of width
$\lambda^2$, which
is verified by numerical calculations~\cite{srgwebsite}.
For momenta within $\lambda$ of the diagonal, the omitted quadratic part of the
flow equation is, of course, essential, and drives the 
flow of physics information necessary to preserve unitarity.

The tool we will use to study decoupling is a
smooth exponential  regulator applied to the potential to
cut off momenta above $\Lambda$:
\beqn
  V_{\lambda,\Lambda}(k,k') =  
  e^{-({k^2}/{\Lambda^2})^{\nexp}} V_{\lambda}(k,k') 
  e^{-({k'^2}/{\Lambda^2})^{\nexp}}  \;,
\label{eq:regulator}
\eeqn
where $\nexp$ takes on integer values. 
From the cut potentials we calculate observables such as phase shifts
and  ground-state energies and 
compare to values calculated with the corresponding uncut potential.
If there is decoupling between
matrix elements in a  given potential (evolved or otherwise), we should be able
to set those elements to zero in this systematic way and use
the relative error in the observable as a metric of the degree
of decoupling. 
By varying $\nexp$ we can identify quantitatively the residual
coupling strength.

The SRG transformations used here are truncated at the two-body level,
which means that they are only approximately unitary for $A > 3$ and
observables will vary with $\flow$.
In these cases decoupling is tested by comparing cut to uncut potentials
at a fixed $\flow$.
All two-body
observables calculated with
the uncut $H_\flow$  are independent of $\flow$ to within numerical precision. 
The actual numerical error depends on the details of the discretization (e.g.,
the number and distribution of mesh points, usually gaussian) and
on the accuracy and tolerances of the
differential equation solver.
While in practice we can make such errors very small, to avoid
mixing up small errors we will also compare cut to uncut potentials rather than
to the unevolved ($s=0$) potential for two-body observables.

\section{Phase Shift Errors}
\label{sec:phase_shifts}

\begin{figure}[ht]
\begin{center}
\includegraphics*[width=6.7cm]{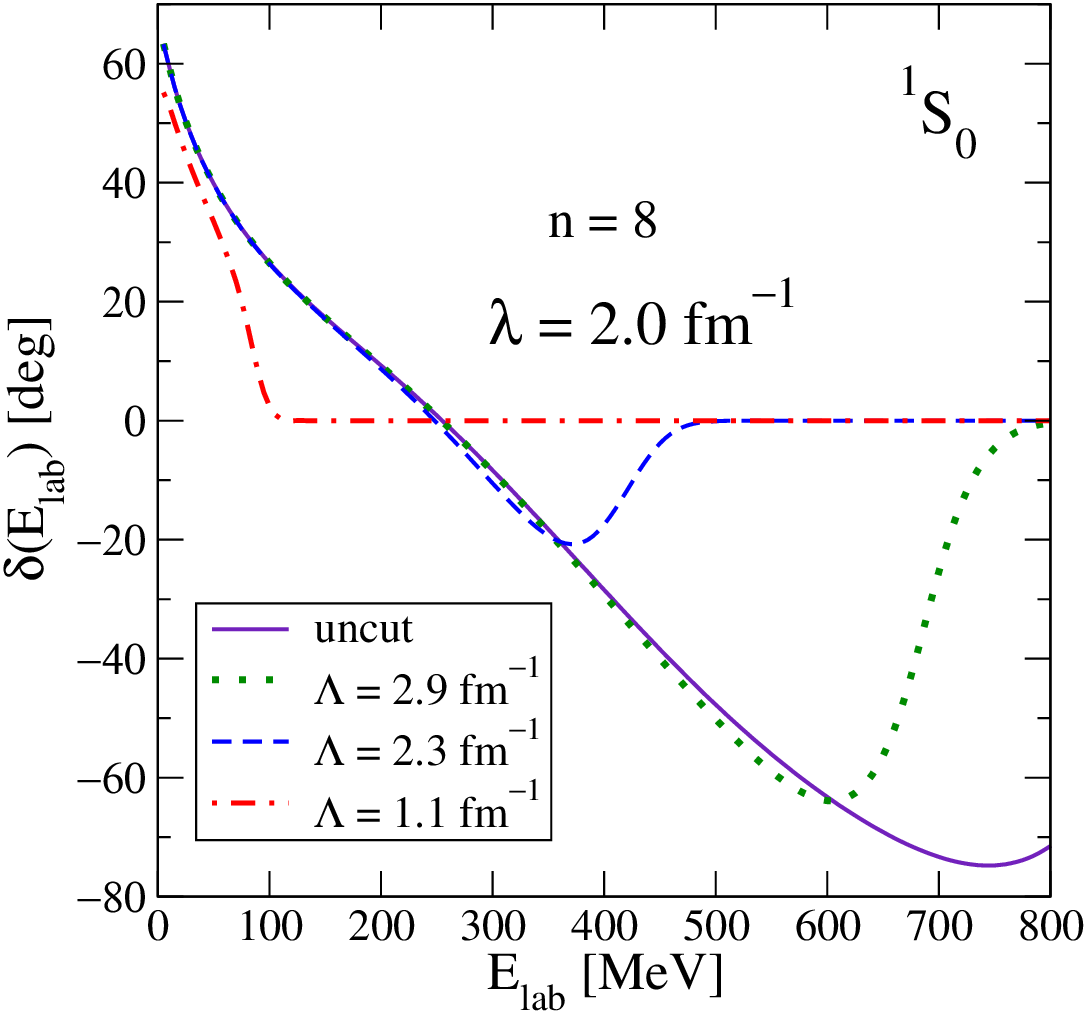}
\hspace*{.1in}
\includegraphics*[width=6.5cm]{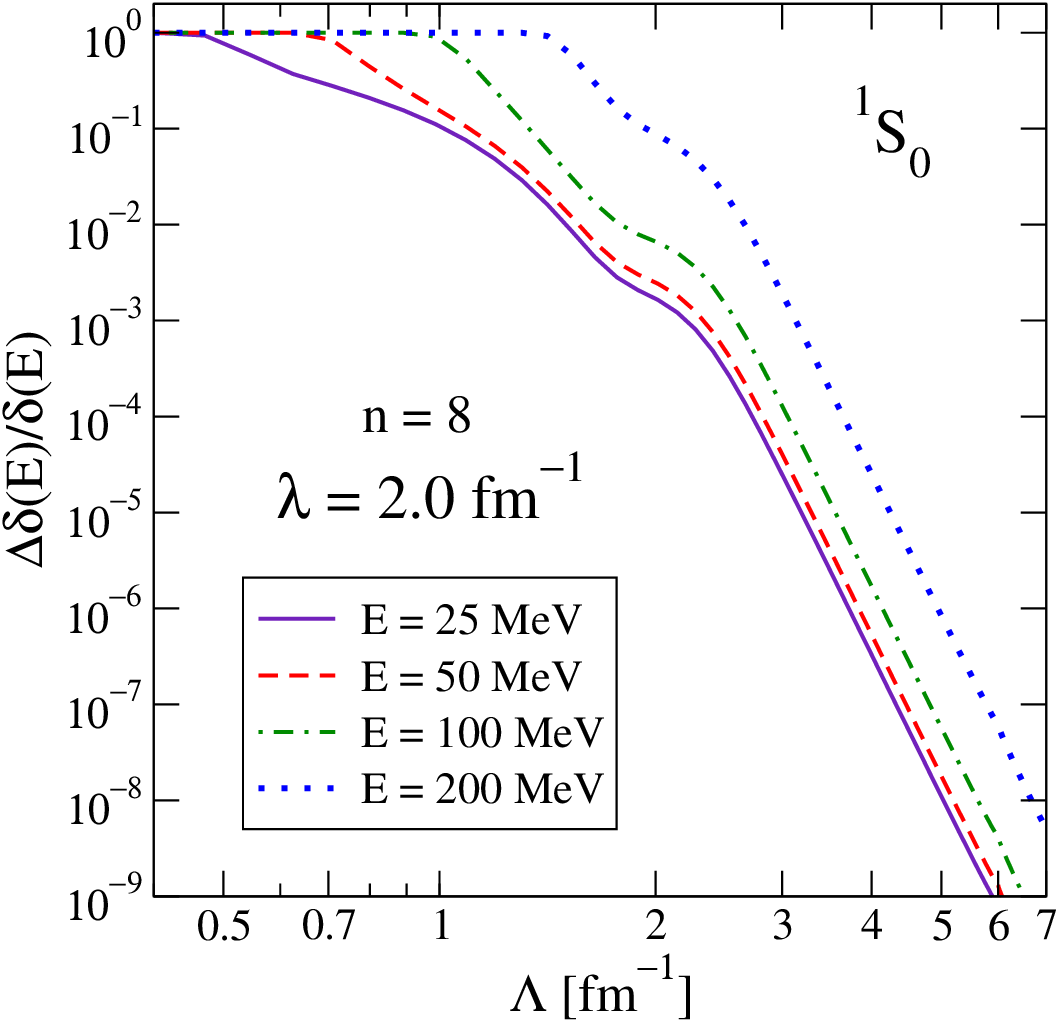}  

\vspace*{.1in}

\includegraphics*[width=6.7cm]{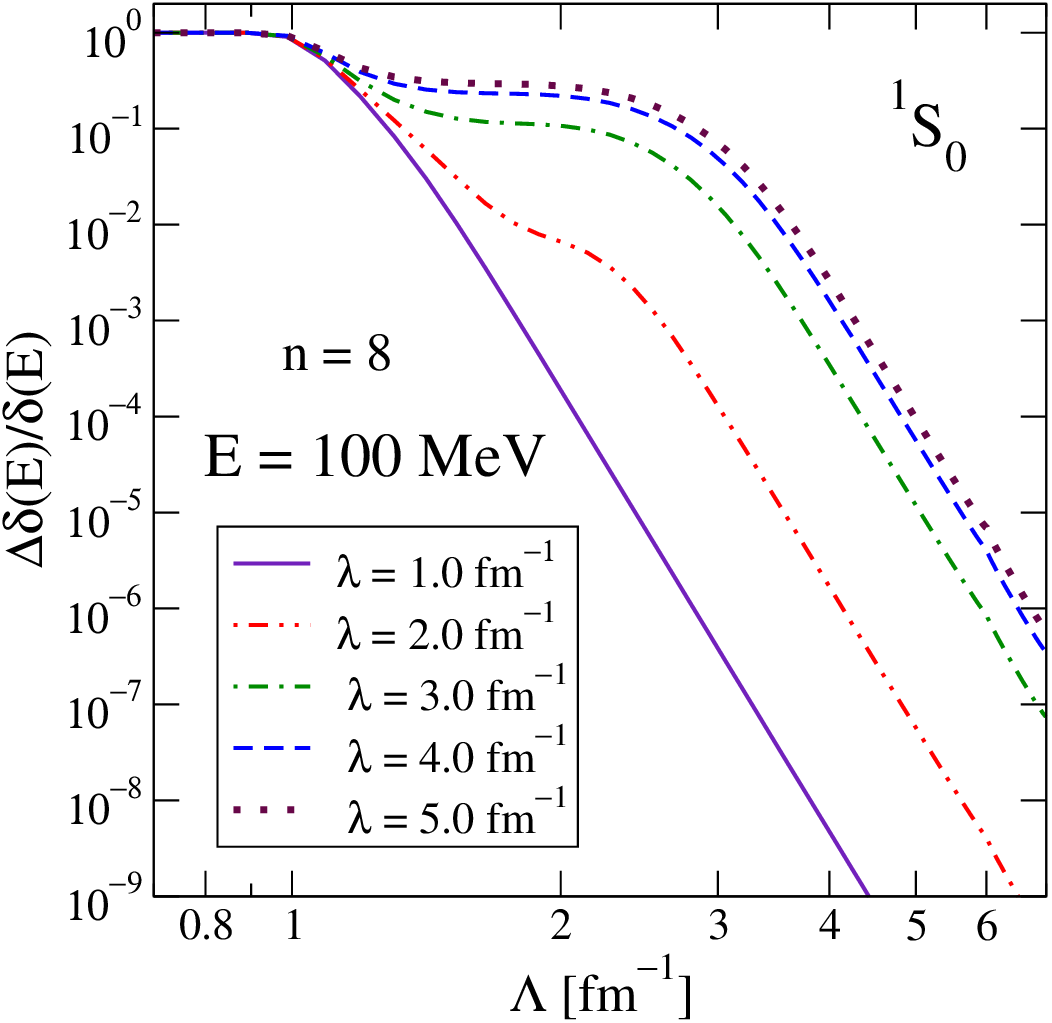}
\hspace*{.1in}
\includegraphics*[width=6.7cm]{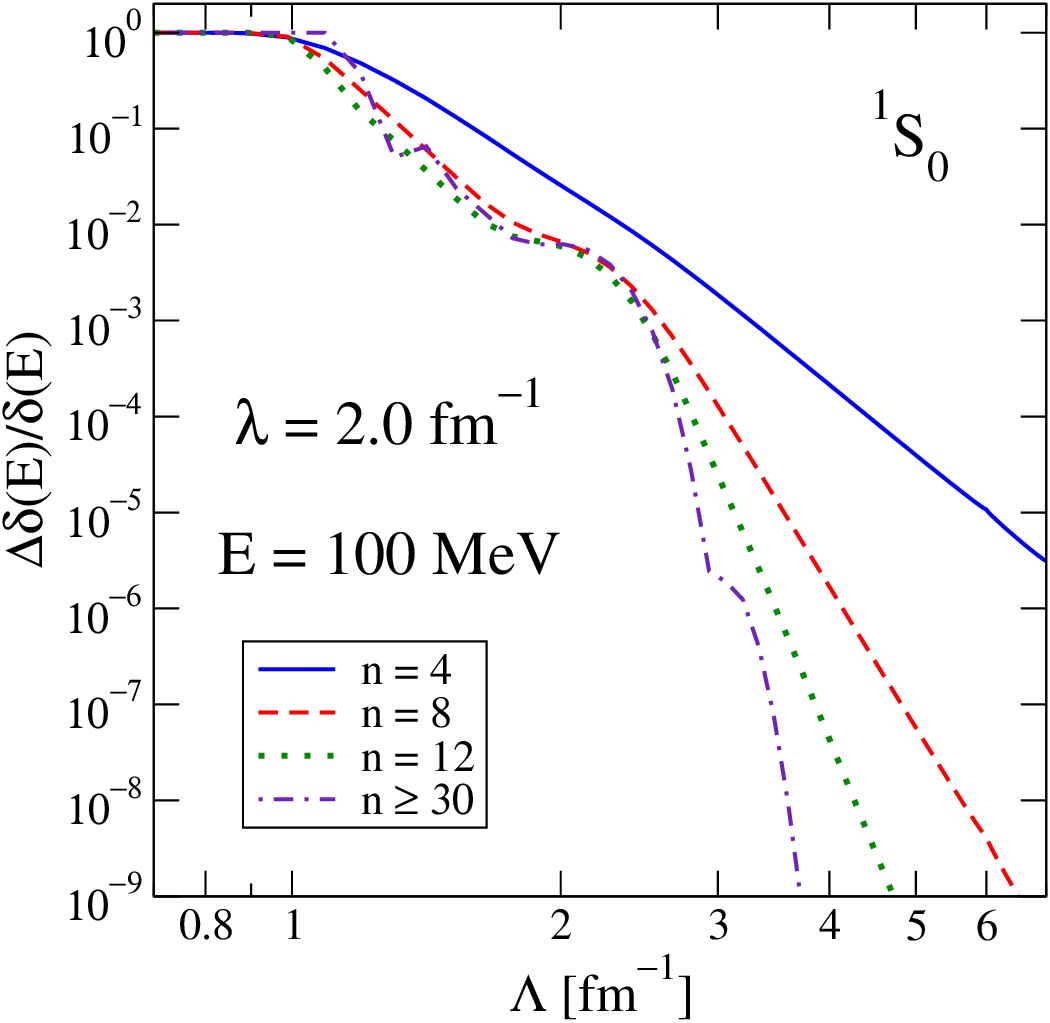}
\end{center}
\caption{Phase shifts and relative errors in the $^1$S$_0$ channel for
SRG potentials evolved from the N$^3$LO (500\,MeV) potential of
Ref.~\cite{N3LO}. The upper-left 
graph shows the phase shifts vs.\ energy for the uncut $\lambda = 2\fmi$
potential and several  
cut versions with $n=8$. 
The other panels show the relative error as a function of the momentum 
cut parameter $\Lambda$ 
at various energies $E$, $\lambda$'s, and $n$'s, respectively.}
\label{fig:ps_err_vs_cut}
\end{figure}

In the upper-left panel of Fig.~\ref{fig:ps_err_vs_cut}, we show 
results for the $^1$S$_0$ phase shifts vs.\ energy calculated using 
the unevolved 500\,MeV N$^3$LO potential of Ref.~\cite{N3LO} and the 
corresponding SRG potential evolved to $\lambda = 2.0\fmi$ and then 
cut using the regulator of Eq.~(\ref{eq:regulator}) with $n=8$. We 
do not explicitly show results from uncut SRG potentials, because 
they are indistinguishable from the unevolved results.

The qualitative pattern is that when the regulator parameter
$\Lambda$ is greater than $\lambda$, there is good agreement of phase shifts
from uncut
and cut potentials at small energies and reasonable agreement up to 
the energy corresponding to the momentum of the cut,
$E_{\rm lab} \approx 2\Lambda^2/m$ (with $\hbar = 1$).
When $V_{\rm srg}$ is cut below $\lambda$, there is poor agreement
everywhere and the phase shift is zero above this energy 
(e.g., above $E_{\rm lab} = 100\,$MeV
for $\Lambda = 1.1\fmi$).  
Thus the decoupling of high and low momentum
means that we can explicitly cut out the high-momentum part of the
evolved potential without significantly distorting low-energy phase shifts 
as long as
we don't cut below $\lambda$.
Cutting out the high-momentum part of conventional nuclear potentials
\textit{does} cause distortions, which
has led to the misconception that reproducing high-energy phase shifts
is important for low-energy nuclear structure observables~\cite{Bogner:2007srg}.

The quantitative
systematics of SRG decoupling is documented in the other panels
of Fig.~\ref{fig:ps_err_vs_cut}, where we look at the relative
error as a function of the cutting momentum using log-log plots.
In these error plots, three main regions are evident. 
In the region below the $\Lambda$ corresponding to the fixed energy,
the predicted phase shift goes to zero since the potential has vanishing
matrix elements, so that the relative error goes to one.
Starting at $\Lambda$ slightly above the value of $\lambda$, there
is a clear power-law decrease in the error.
In between is a transition region without a definite pattern.

We focus here on the power-law region. In the lower-left pane, we find 
that this decoupling starts with a shoulder at momenta slightly above 
$\lambda$. This effect saturates when $\lambda$ becomes comparable to 
the underlying cutoff of the original potential. In the upper-right pane 
we see that the shoulder signaling the start of the power-law decrease is 
not affected by the energy, $E$. This holds for other values of $\lambda$ 
and $n$. In the lower-right pane we vary  the exponent of the regulator, 
$n$, which changes the smoothness of the regulator. The smoothness affects 
the slope of the power law and the fine details in the intermediate region, 
but does not change the  position of the shoulder near $\lambda$. As discussed 
below, the power-law behavior in the  relative error signifies perturbative 
decoupling with a strength given by the sharpness of the regulator used to 
cut off the potential.

\begin{figure}[ht]
\begin{center}
\includegraphics*[width=6.5cm, clip=true]{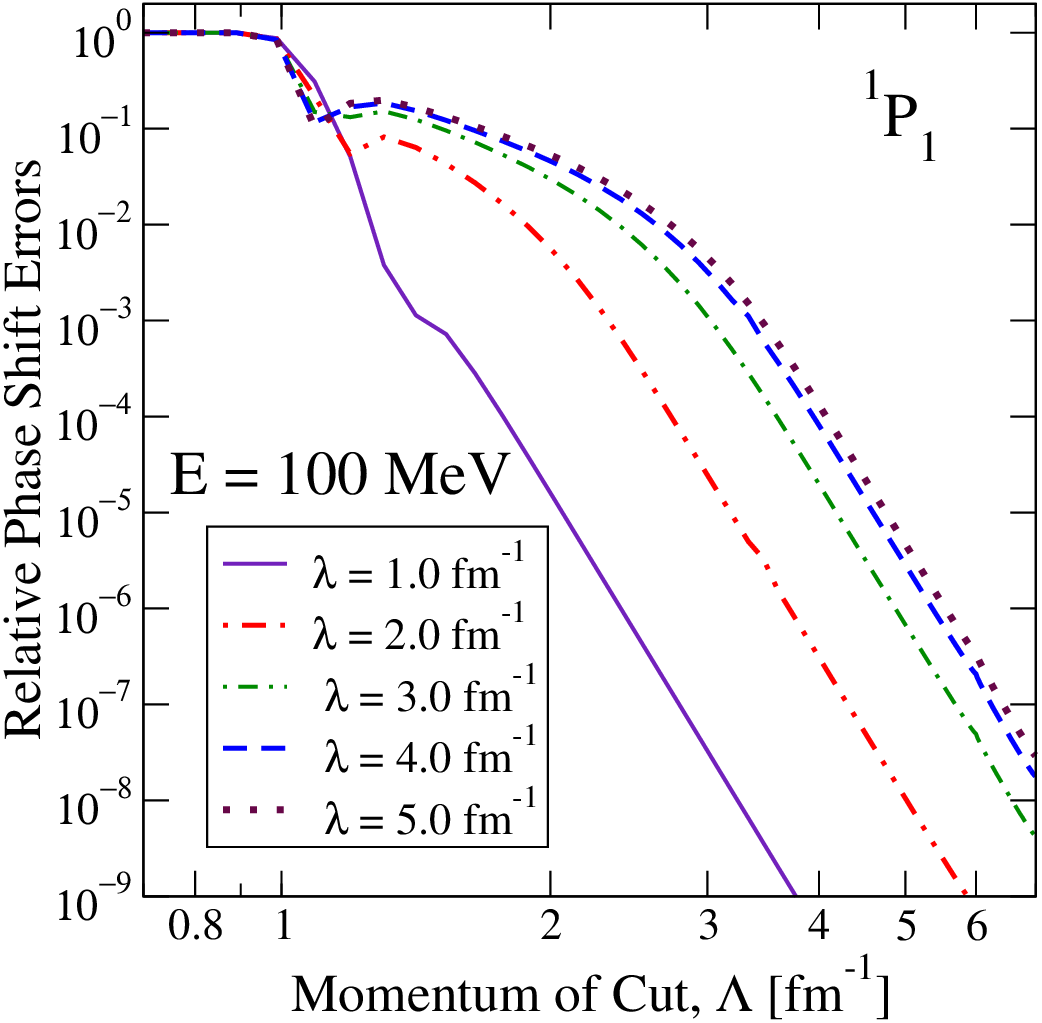} 
\includegraphics*[width=6.5cm, clip=true]{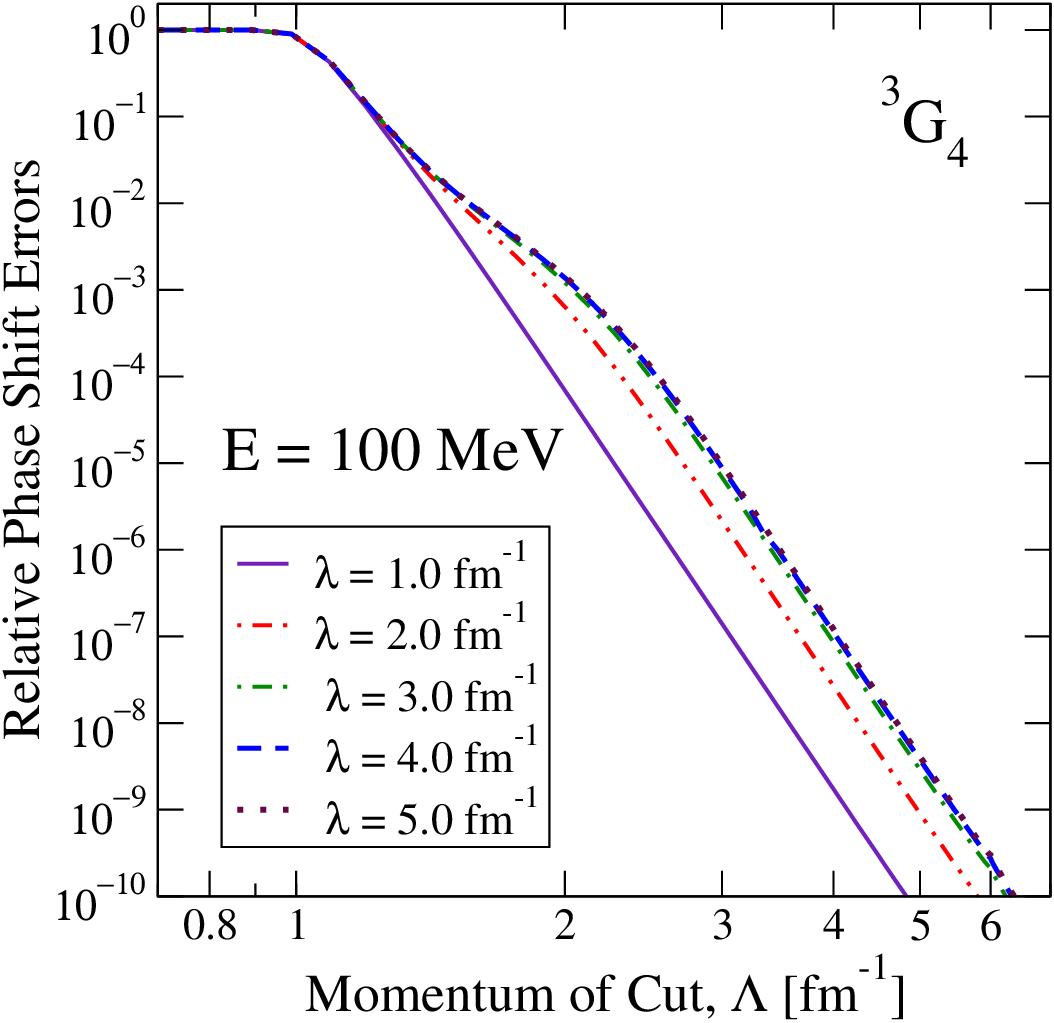}
\end{center}
\caption{The phase shift errors computed in select partial waves, using
a regulator with $n=8$. 
Other channels exhibit the same power-law dependence of the error
for $\Lambda > \lambda$.}
\label{fig:ps_err_other_channels}
\end{figure}

We checked this decoupling behavior in different partial waves
and for other N$^3$LO potentials and found 
the same perturbative region in all cases. 
Representative partial waves are shown for 
various $\lambda$ values in Fig.~\ref{fig:ps_err_other_channels}. 
The potential in the S waves typically passes
through zero for momenta in the region of $\lambda = 2\fmi$
(see, for example, Ref.~\cite{srgwebsite}),
which might lead one to associate decoupling with 
this structure.
The error plots for other partial waves that lack this structure
show that it is a more general consequence of the SRG evolution.

Indeed,
the behavior of the errors in the decoupling region, where $\Lambda > \lambda$,
can be directly understood as a consequence of the partial diagonalization
of the evolved potential.
The calculation of the phase shift at a low-energy $k^2 \ll \lambda^2$
will involve an integral over $p$ of $V_{\lambda,\Lambda}(k,p)$.
But the potential cuts off the integral at roughly 
$p^2 \approx k^2 + \lambda^2 < \Lambda^2$, 
which means that we can expand the difference in the uncut
and cut potentials:
\beqn
 \delta V_{\lambda,\Lambda}(k,p) \equiv
    V_{\lambda}(k,p) - V_{\lambda,\Lambda}(k,p)
    \approx \left(\frac{k^{2n}}{\Lambda^{2n}}
                 + \frac{p^{2n}}{\Lambda^{2n}}  \right) V(k,p)
    \;. 
    \label{eq:deltaV}
\eeqn
Simple perturbation theory in $\delta V$
then predicts the dependence of the phase shift error to be $1/\Lambda^{2n}$,
which is the power-law dependence seen in Figs.~\ref{fig:ps_err_vs_cut} and
\ref{fig:ps_err_other_channels}.
The accuracy of first-order perturbation theory is evidenced by the 
constant slope of the error curves, which translates into perturbatively
small residual coupling.

The detailed dependence on the energy and $\lambda$ is not so trivially
extracted.
However,
the weak dependence on $E_{\rm lab} \leq 100\,\mbox{MeV}$ and strong
dependence on $\lambda < 3 \fmi$ at fixed $\Lambda$ seen in 
Fig.~\ref{fig:ps_err_vs_cut} implies that the integration picks up
the scale $\lambda$, so that the dominant
error scales as $(\lambda/\Lambda)^{2n}$.
This is, in fact, observed numerically for intermediate values of
$\lambda$ (e.g., for $1.8\fmi < \lambda < 2.8\fmi$ when $\Lambda = 3\fmi$). 

\section{Decoupling and Deuteron Observables}
\label{sec:deuteron}

\begin{figure}[ht]
\begin{center}
\includegraphics*[width=6.5cm, clip=true]{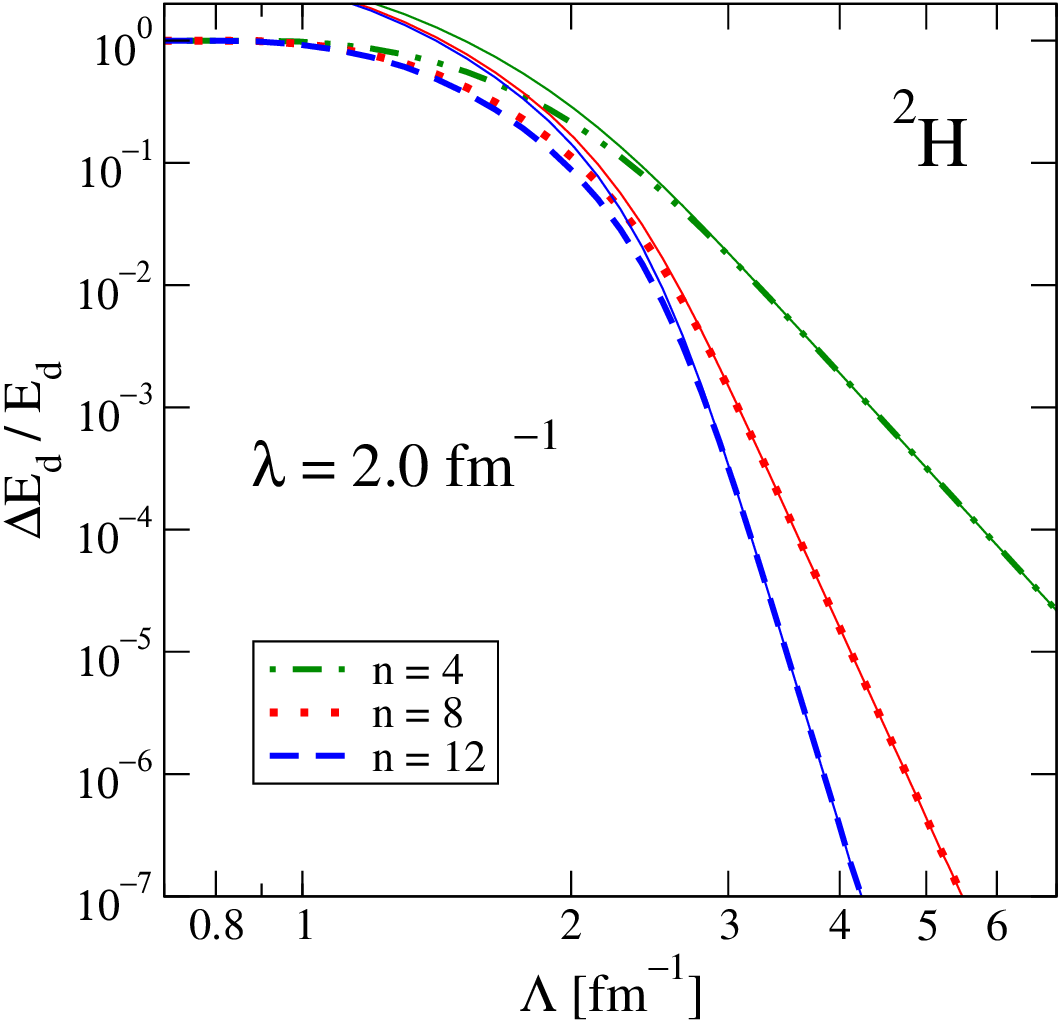} 
\hspace*{.1in}
\includegraphics*[width=6.5cm, clip=true]{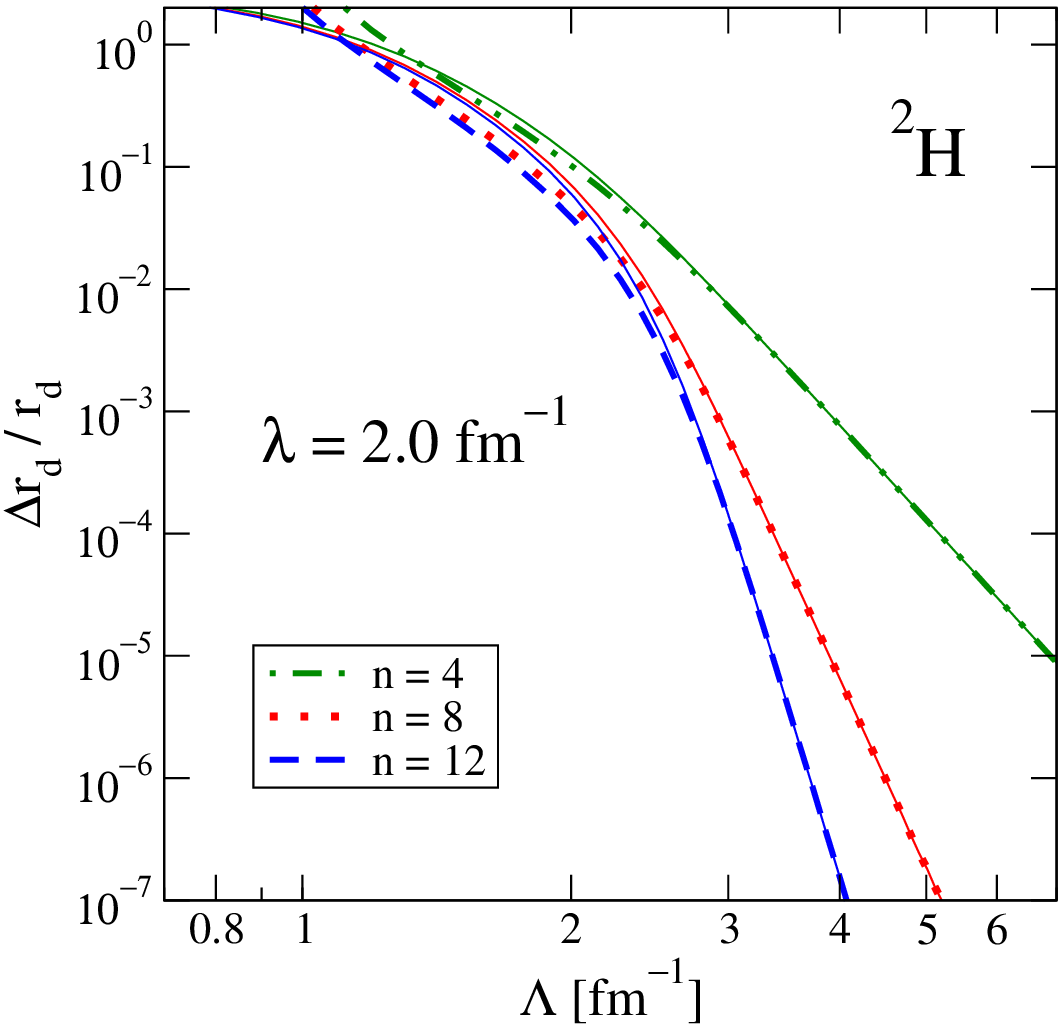} 
\end{center}
\caption{The relative error vs. cut parameter $\Lambda$ of 
the deuteron energy (left) and rms radius (right)
with several values of the regulator 
parameter $n$ indicated in the legends. 
In each case,
the near-by solid line is the estimate of the error using first-order
perturbation theory with Eq.~(\ref{eq:deltaV}).}
\label{fig:deut_err_vs_cut}
\end{figure}

To test the generality of the observations made for phase shifts,  the
same techniques were applied to other low-energy observables such as
the deuteron binding energy, radius, and quadrupole moment. 
The binding energy and momentum-space wavefunction 
were computed using standard
eigenvalue methods. The computation of $Q_d$ and $r_d$
from the wavefunction uses~\cite{Bogner:2006vp},

\bea
Q_d = -\frac{1}{20} \int_0^\infty \! dk \, \biggl[
\sqrt{8} \, \biggl( k^2\, \frac{d\wt u(k)}{dk} \frac{d\wt w(k)}{dk}
+ 3 k\, \wt w(k) \frac{d\wt u(k)}{dk} \biggr) \nonumber \\
 + k^2 \biggl( \frac{d\wt w(k)}{dk}\biggr)^2 + 6 \, \wt w(k)^2 \biggr] \;, 
\label{eq:quad}
\eea
and
\beqn
r_d = \frac{1}{2} \biggl[ \int_0^\infty \! dk \, \biggl\{ \biggl(
k \, \frac{d\wt u(k)}{dk} \biggr)^2 + \biggl(k \, \frac{d\wt w(k)}{dk}\biggr)^2 
+ 6 \, \wt w(k)^2 \biggr\} \biggr]^{1/2} \;,
\label{eq:rad}
\eeqn

where $\wt u(k$) and $\wt w(k)$ correspond to the S and D components 
of the deuteron wavefunction respectively. We again computed relative 
errors in these observables and, as shown in Fig.~\ref{fig:deut_err_vs_cut}
for the energy and radius, 
the errors show the same behavior as observed for
the phase shifts.
That is, a power-law drop-off in the error begins at $\Lambda$ just above 
$\lambda$, with a slope determined by the sharpness of the regulator
as given by $n$. 
The relative error for the
quadrupole moment is not shown but is very similar to that for the radius.

As with the phase shift, the analytic dependence of the
error from cutting the potential can be estimated directly
in perturbation theory. 
In this case, partial diagonalization of the potential means
that the deuteron wave function has negligible momentum components
starting slightly above $\lambda$.
This in turn validates the expansion in Eq.~(\ref{eq:deltaV})
and the dependence of the errors on $1/\Lambda^{2n}$.
The numerical calculation of the error in perturbation theory is
plotted in Fig.~\ref{fig:deut_err_vs_cut} and shows close agreement
in the decoupling region $\Lambda > \lambda$.

\section{Decoupling and Few-Body Energies with the NCSM}
\label{sec:ncsm_calcs}

The calculations described above have been only for two-particle systems.
Using NCSM calculations of ground-state energies with
the Many-Fermion Dynamics (MFD) code~\cite{MFD}, we can test
whether the high-energy decoupling behavior extends to few-body 
systems.
In the present study,
only NN interactions were considered, with
the testing of decoupling with many-body forces deferred
to a future investigation.
However, we note that the general
features of the SRG exhibited in Section~\ref{sec:methods}
implies that off-diagonal matrix elements (with respect to energy) of
the three-body force will be suppressed, with decoupling as an
expected consequence.

\begin{figure}[ht]
\begin{center}
\includegraphics*[width=6.5cm]{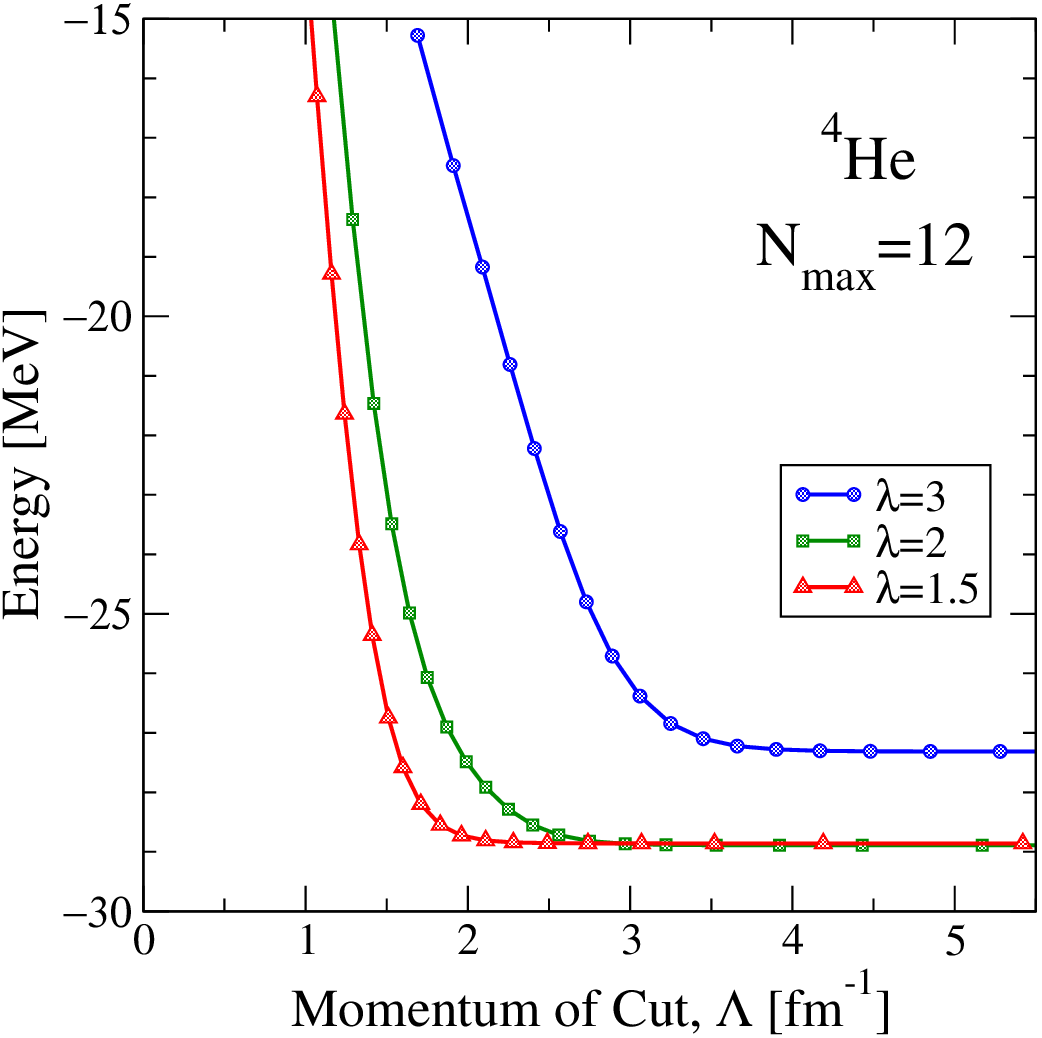} 
\quad
\includegraphics*[width=6.5cm]{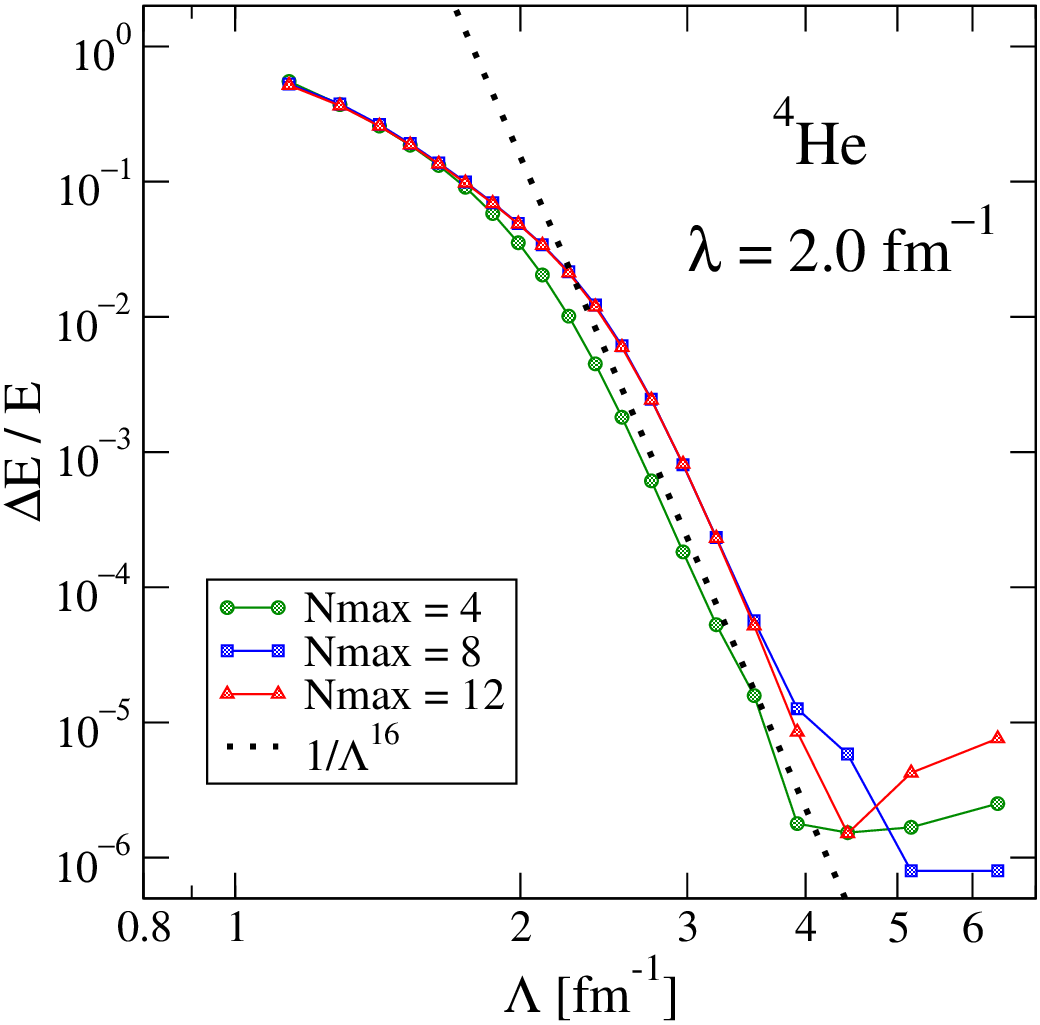}
\end{center}
\caption{Calculations of the $^4\rm{He}$ ground-state energy
using the NCSM. On the left is 
the energy obtained from the NCSM for potentials evolved to several
different $\lambda$ values as a function of the cut (regulator) momentum
$\Lambda$ with $n=8$.  On the right is the relative 
error of the energy
for the $\lambda = 2\fmi$ case as a function of the cut momentum
(with $n=8$)
for several different harmonic oscillator basis sizes.
Also shown is the slope of the error in the decoupling region
predicted from perturbation theory (dotted line).}
\label{fig:ncsm_nuclei_err_vs_cut_He4}
\end{figure}

\begin{figure}[ht]
\begin{center}
\includegraphics*[width=6.5cm]{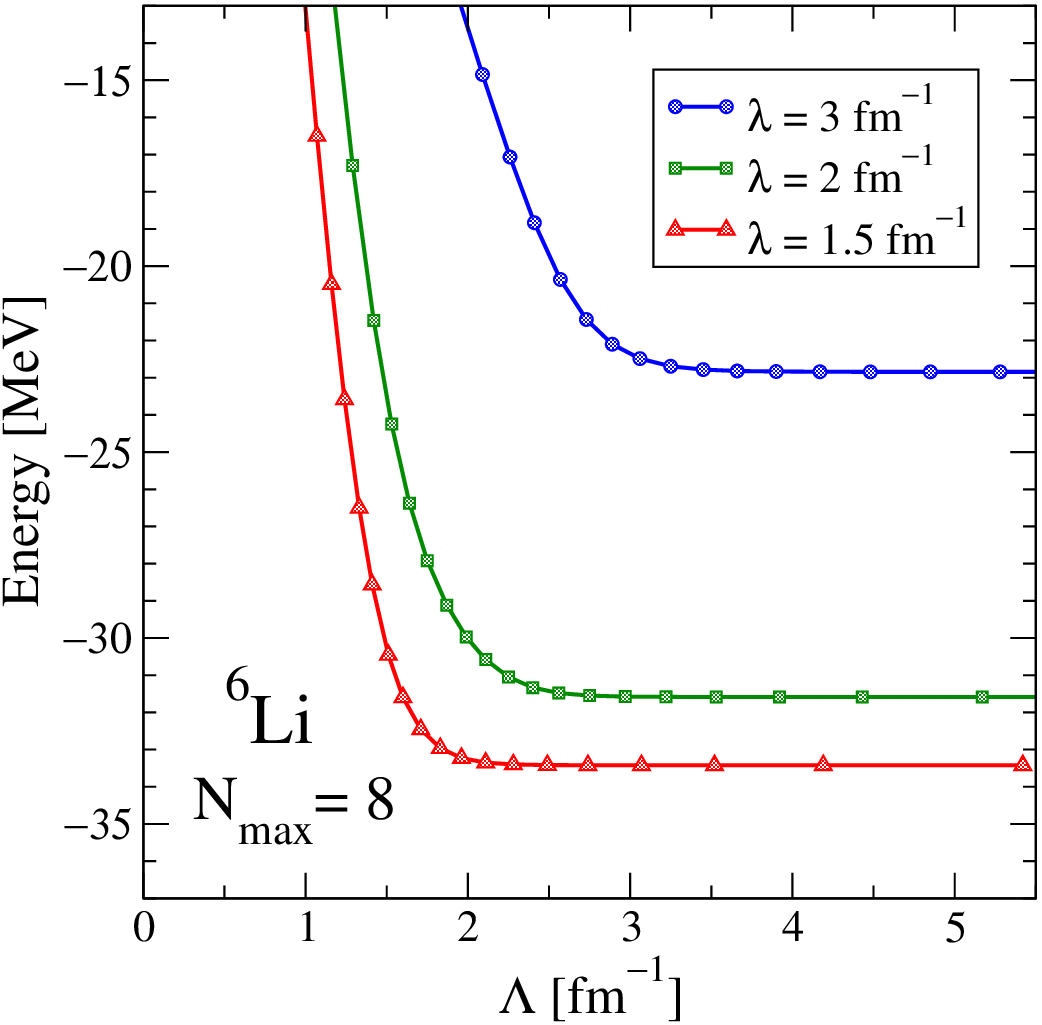} 
\quad
\includegraphics*[width=6.5cm]{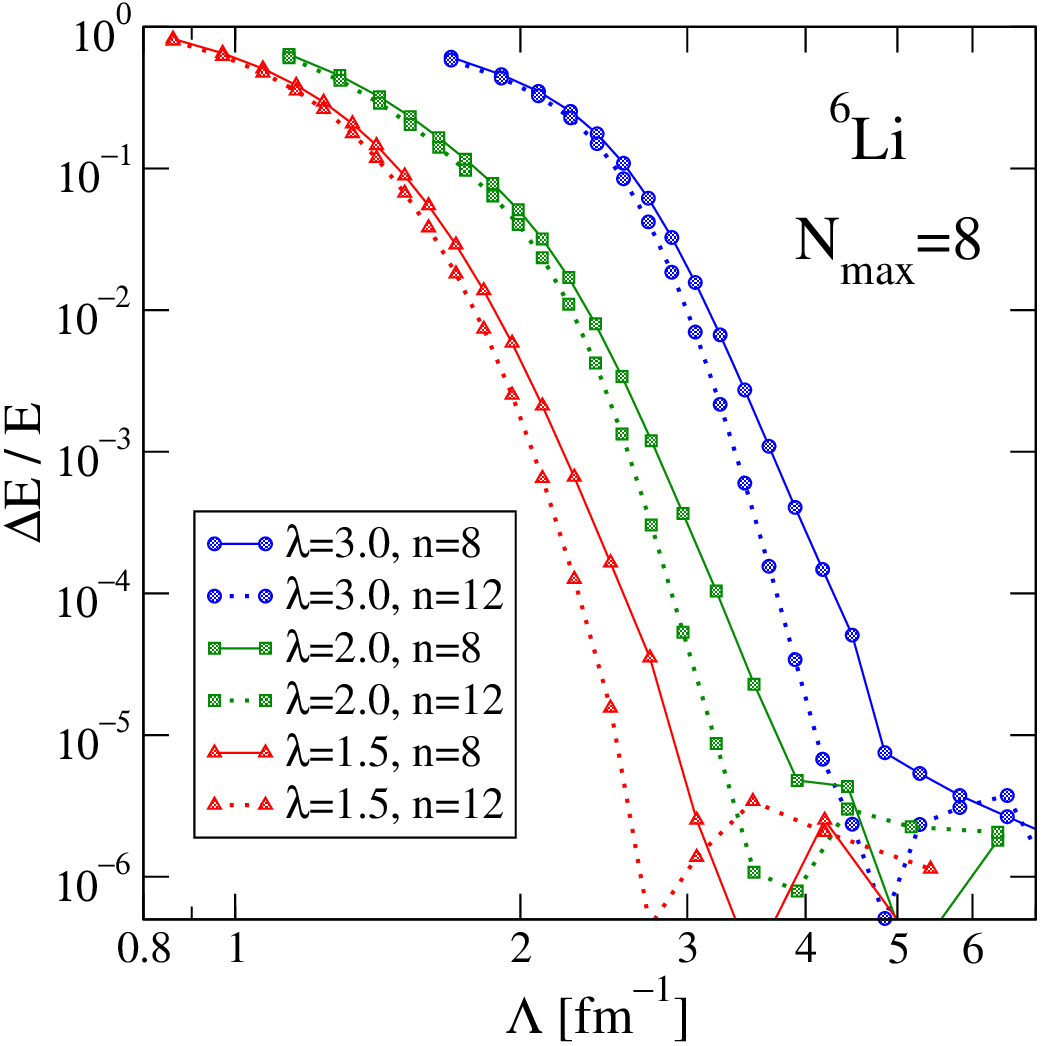}
\end{center}
\caption{Calculations of the $^6\rm{Li}$ ground-state energy
using the NCSM. On the left is 
the energy obtained from the NCSM for potentials evolved to several
different $\lambda$ values as a function of the cut (regulator) momentum
$\Lambda$ with $n=8$.  On the right is the relative 
error of the energy
for the same $\lambda$'s as a function of the cut momentum
for the same $\lambda$ values but with two values of $n$.}
\label{fig:ncsm_nuclei_err_vs_cut}
\end{figure}

We first verified that the decoupling behavior already
observed using a direct calculation of the deuteron
wavefunction is reproduced using the MFD. 
We then calculated a series of 
larger nuclei, including $^3$H, $^4$He, and $^6$Li,
comparing results from uncut and a range of cut potentials evolved
to different values of $\lambda$.
On the left panel of Fig.~\ref{fig:ncsm_nuclei_err_vs_cut_He4},
the $^4$He ground-state energy is plotted versus 
the regulator parameter $\Lambda$ 
for several different values of the SRG flow parameter $\lambda$. 
Each of the plotted points is at a basis size $N_{\rm max} = 12$;
this is within several hundred keV of the energy from
extrapolating to $N_{\rm max} = \infty$.
A similar plot for $^6$Li is given in the left panel of 
Fig.~\ref{fig:ncsm_nuclei_err_vs_cut} using 
a basis size $N_{\rm max} = 8$,
which is within several hundred keV of the extrapolated energy
for $\lambda = 1.5\fmi$ but still several MeV off for 
$\lambda = 3.0\fmi$.
The uncut ($\Lambda \rightarrow \infty$) energies vary for each
$\lambda$ because the SRG evolution includes the NN interaction
only; the closeness of the results for $\lambda = 2\fmi$ and $3\fmi$
for $^4$He is accidential (see Ref.~\cite{Bogner:2007rx} for further
discussion about the running of the energies).

Both examples show that
when the potential is cut with $\Lambda$ comparable to $\lambda$ or lower,
the converged energy is significantly different from the asymptotic uncut
value, while it approaches that value rapidly as $\Lambda$ moves above
$\lambda$. 
This means that, for smaller $\lambda$, more high 
momentum matrix elements can be discarded without a loss of accuracy. 
This decoupling explains 
the greatly improved convergence with basis size seen in the 
NCSM for corresponding
$\lambda$ values~\cite{Bogner:2007rx}.

The quantitative behavior of the relative error parallels that observed for
two-body observables, as seen 
on the right panels of Figs.~\ref{fig:ncsm_nuclei_err_vs_cut_He4}
and \ref{fig:ncsm_nuclei_err_vs_cut}.
In all cases, for a fixed value of $\lambda$ the power decrease
in the error starting with $\Lambda$
slightly above $\lambda$ is clearly seen,
even though there are  fewer digits of precision in the NCSM 
results (so the relative error is in the range $10^{-6}$--$10^{-5}$ at best).
The same perturbative residual coupling is seen for different basis
sizes, with the slope given by the
dependence $1/\Lambda^{2n}$, although the onset of the decoupling
region shifts to higher $\Lambda$ until the calculation is near
convergence (see Fig.~\ref{fig:ncsm_nuclei_err_vs_cut_He4}). 
Similar results are found for other nuclei and for other values
of $\lambda$.

We repeated the calculations 
with other choices of the SRG generator, $G_\flow$
including $\Hzero^2$ and 
$H_{\rm D} = \Hzero + V_{\rm D}$, where $V_{\rm D}$ is the 
(running) diagonal part of the bare potential. 
We found that these other 
choices for $G_\flow$ do not alter the power-law 
behavior region of the previous error plots. 
This provides further 
evidence that the high-  and low-energy decoupling results
primarily from 
the partially diagonalized nature of the evolved potential.

\section{Summary}
\label{sec:summary}

The evolution of nucleon-nucleon
potentials with the Similarity Renormalization Group decouples 
high-energy degrees of freedom from calculations
of low-energy observables. 
By using 
a steep but smooth exponential function of the 
form $\exp[-(k^2/\Lambda^2)^{\nexp}]$ with integer $\nexp$
to set interaction
matrix elements with relative momenta above $\Lambda$ smoothly
to zero, 
we have found that the residual 
coupling follows a clear and universal behavior.
Decoupling is achieved for $\Lambda$ above $\lambda$, the flow
parameter for the SRG.
The dependence on $\nexp$
was used to study the region of weak residual coupling, which was
found to follow a power law predicted 
by leading-order perturbation theory.
These results were shown to 
apply to NN phase shifts, several deuteron observables, and the
ground-state energy of nuclei up to $A = 6$.
Similar results are also found for other generators that are diagonal
in momentum space.

We emphasize that the regulator used here was only a tool for
studying the effect of the SRG on the potential. 
No cuts of this nature have been used or
are proposed for calculations with NN potentials.
However, the most important next step with the SRG is
the evolution of three-body forces~\cite{Bogner:2007qb}, which 
have significantly larger computational requirements.
There is every indication that the SRG will induce the same
decoupling for interactions that include three-body forces. 
These potentials will
be projected on a momentum basis ordered by the kinetic energy
of the states, and the SRG evolution equations will suppress
matrix elements that are off-diagonal in that energy basis~\cite{Bogner:2007qb}.
Decoupling as observed here would
allow a three-body
computation to freeze irrelevant high-energy details during the evolution
in a controlled way, which could significantly reduce the computational
resources needed.

\begin{ack}
This work was supported in part by the National Science Foundation
under Grant Grant Nos.~PHY--0354916 and PHY--0653312, 
and the the UNEDF SciDAC Collaboration under DOE Grant 
DE-FC02-07ER41457.
\end{ack}


\begin{thebibliography}{99} 

\bibitem{Glazek:1993rc} S.D. Glazek and K.G. Wilson, Phys. Rev. 
  D \textbf{48} (1993) 5863; Phys. Rev. D \textbf{49} (1994) 4214.

\bibitem{Wegner:1994} F. Wegner, Ann. Phys. (Leipzig) \textbf{3} (1994) 77;
  Phys.\ Rep.\ {\bf 348} (2001) 77.

\bibitem{Kehrein:2006}
J. Kehrein, {\it The Flow Equation Approach to Many-Particle Systems\/}
    (Springer, Berlin, 2006).

\bibitem{Bogner:2006srg} S.K. Bogner, R.J. Furnstahl, and R.J. Perry,
Phys. Rev. C {\bf 75}  (2007) 061001.

\bibitem{Bogner:2007srg}S.K. Bogner, R.J. Furnstahl, R.J. Perry, and A. Schwenk,
Phys. Lett. B {\bf 649} (2007) 488.

\bibitem{Wiringa:1994wb} R.B. Wiringa, V.G.J. Stoks and R. Schiavilla,
Phys. Rev. C \textbf{51} (1995) 38.

\bibitem{N3LO} D.R. Entem and R. Machleidt,  
 Phys. Rev. C \textbf{68} (2003) 041001(R).  

\bibitem{N3LOEGM} E. Epelbaum, W. Gl\"ockle and U.G. Mei{\ss}ner,
Nucl. Phys. \textbf{A747} (2005) 362.
 
\bibitem{Bogner:2007rx}
  S.K. Bogner, R.J. Furnstahl, P. Maris, R.J. Perry, A. Schwenk and J.P. Vary,
  arXiv:0708.3754 [nucl-th].

\bibitem{NCSMC12} P. Navr\'atil, J.P. Vary and B.R. Barrett,
                   Phys.\ Rev.\ Lett.\ {\bf 84} (2000) 5728;
                   Phys.\ Rev.\ C {\bf 62} (2000) 054311.
		   
\bibitem{NCSM2} P. Navr\'atil and W.E.\ Ormand, 
                Phys.\ Rev.\ Lett.\  {\bf 88} (2002) 152502.    

\bibitem{NCSM3}  P. Navr\'atil and W.E.\ Ormand,		   
                Phys.\ Rev.\ C {\bf 68} (2003) 034305. 
  
\bibitem{Nogga:2005hp}
  A.~Nogga, P.~Navr\'atil, B.R.~Barrett and J.P.~Vary,
  Phys.\ Rev.\  C {\bf 73} (2006) 064002.

\bibitem{Navratil:2007we}
  P.~Navr\'atil, V.G.~Gueorguiev, J.P.~Vary, W.E.~Ormand and A.~Nogga,
  Phys.\ Rev.\ Lett.\ {\bf 99} (2007) 042501.
  
\bibitem{Glazek:2007} S.D.\ Glazek and R.J.\ Perry, 
 \textit{The impact of bound states on similarity renormalization group 
 transformations}; in preparation.

\bibitem{srgwebsite} See http://www.physics.ohio-state.edu/$\sim$ntg/srg/ 
for documentary examples.

\bibitem{Bogner:2006vp} S.K. Bogner, R.J. Furnstahl, S. Ramanan and A. Schwenk,
  Nucl. Phys.  A {\bf 784}, 79 (2007).
  
\bibitem{MFD}
 J.P.\ Vary, The Many-Fermion Dynamics Shell-Model Code,
 Iowa State University (1992) (unpublished); J.P.\ Vary and D.C.\ Zheng,
 ibid., (1994) (unpublished).   


\bibitem{Bogner:2007qb}
  S.~K.~Bogner, R.~J.~Furnstahl and R.~J.~Perry,
  arXiv:0708.1602 [nucl-th].

 
\end{thebibliography}
\end{document}